\documentstyle[times,twocolumn,latex8]{article}

\begin{document}

\newtheorem{lemma}{Lemma}
\newtheorem{definition}{Definition}
\newtheorem{theorem}{Theorem}
\newtheorem{proposition}{Proposition}
\newtheorem{corollary}{Corollary}
\newcommand{\debproof}[1]{\par \addvspace{5pt} \noindent {\em Proof{#1}.} }
\newcommand{\finproof}{\mbox{$\hfill \Box$} \par \addvspace{10pt}}

\newcommand{\Ha}{\mbox{$H_A$}}
\newcommand{\Hb}{\mbox{$H_B$}}
\newcommand{\ket}[1]{\mbox{$| #1 \rangle$}}
\newcommand{\bra}[1]{\mbox{$\langle #1 |$}}
\newcommand{\bb}[2]{\mbox{$b_{#1}(#2)$}}
\newcommand{\bbA}[2]{\mbox{$a_{#1}(#2)$}}
\newcommand{\bbB}[2]{\mbox{$b_{#1}(#2)$}}

\title{Quantum Cryptography with Imperfect Apparatus\thanks{This
research was supported in part by DIMACS, and
by DARPA/ITO and the National Science Foundation
under
Grant CCR-9627819.}}

\author{Dominic Mayers\\
Computer Science Department\\
Princeton University\\ Princeton, NJ 08544\\ 
mayers@cs.princeton.edu\\
\and
Andrew Yao\\
Computer Science Department\\
Princeton University\\ Princeton, NJ 08544\\ 
yao@cs.princeton.edu\\
}

\maketitle
\thispagestyle{empty}

\begin{abstract}
Quantum key distribution, first proposed by Bennett and
Brassard, provides a possible key distribution scheme
whose security depends only on the quantum laws of
physics.  So far the protocol has been proved
secure even under channel noise and detector faults of
the receiver, but is vulnerable if the photon source
used is imperfect.  In this paper we propose
and give a concrete design for a
new concept, {\it self-checking source}, which
requires the manufacturer of the photon source
to provide certain tests; these tests are designed
such that, if passed, the source is guaranteed to
be adequate for the security of the quantum key
distribution protocol, even though the testing
devices may not be built to the original specification.
The main mathematical result is a structural
theorem which states that, for any
state in a Hilbert space, if certain
EPR-type equations are satisfied, the state
must be essentially the orthogonal sum of
EPR pairs.
\end{abstract}

\section{Introduction}
In 1984, Bennett and Brassard \cite{bb84} proposed a revolutionary
concept that {\it key distribution} may be accomplished through public
communications in quantum channels.  Hopefully, the privacy of the
resulted key is to be guaranteed by quantum physical laws alone, quite
independent of how much computational resource is available to the
adversary.  The primary quantum phase of the proposed protocol is a
sequence of single photons produced by Alice (the sender) and detected
by Bob (the receiver).

The security proof of the BB84-protocol (or its many variants) for
adversaries with unrestricted power is a difficult mathematical
problem, and has only been achieved with any generality in the last
few years.  In brief, the BB84-protocol is secure even with channel
noise and possible detector faults for Bob, provided that the
apparatus used by Alice to produce the photons is perfect.  The
purpose of this paper is to remove this last assumption, by proposing
and giving a concrete design for a new concept, {\it self-checking
source}, which requires the manufacturer of the photon source to
provide certain tests; these tests are designed such that, if passed,
the source is guaranteed to be adequate for the security of the
BB84-protocol, even though the testing devices may not be built to the
original specification.  A self-checking source must receive inputs
from multiple locations (two in our case) and returns classical
outcomes at these locations. The test needs only to consider the
classical inputs and the classical outcomes.

It is well known that there are clever ways to construct imperfect
sources for the coding used in the BB84-protocol that behave quite
normal on the surface, but seriously compromise the security. In other
words, the BB84 coding together with the standard test executed in the
BB84-protcol are problematic because the external data can be
reproduced by quantum apparatus which are not secure at all.  We
propose a different source that is self-checking and yet can be used
to generate the BB84 coding.  Our result means that one does not have
to perform an infinite number of ways to check all possible devious
constructions. In some ways our test can be regarded as simple
self-testing quantum programs.  Our result requires that, when
the inputs to the source are fixed, the distribution of probability
for the classical outcomes is also fixed.

Our result is that, if these distributions of probability (associated
with the different inputs) are exactly as in the specification for our
self-checking source, the state transmitted is a direct sum of states
that are individually normally emitted by a perfect source.  In
practice, we cannot expect these probabilities to be exactly as in the
specification for the self-checking source. However, one can test that
they are not too far away from this specification.  Furthermore, one
should expect that the closer to their specified values these
probabilities will be, the closer to the direct sum described above
the source will be.  This is usually sufficient to prove security.

In Section 2, we show how the main mathematical question arises from
the security requirement from the BB84-protocol.  In Section 3, the
precise question is formulated, and the main theorem stated.  The
proof of the main theorem is given in Section 4.

\section{Preliminaries}
Ideally, the objective of key distribution is to allow two
participants, typically called Alice and Bob, who initially share no
information, to share a secret random key (a string of bits) at the
end.  A third party, usually called Eve, should not be able to obtain
any information about the key.  In reality, this ideal objective
cannot be realized, especially if we give unlimited power to the
cheater, but a quantum protocol can achieve something close to it.
See \cite{bbbss92} (and more recently \cite{mayers98}) for a detailed
specification of the quantum key distribution task.  One of the
greatest challenges in quantum cryptography is to prove that a quantum
protocol accomplishes the specified task.  One can experimentally try
different kinds of attacks, but one can never know in which way the
quantum apparatus can be defective.  In any case, such experiments are
almost never done in practice because it is not the way to establish
the security of quantum key distribution.  The correct way is a
properly designed protocol together with a security proof.

Recently, there has been a growing interest in practical quantum
cryptography and systems have been implemented
\cite{trt93b,mbg93,franson94,hughes96,jf96,bhklmnp98}.  However,
proving the security of quantum key distribution against {\em all}
attacks turned out to be a serious challenge.  During many years, many
researchers directly or indirectly worked on this problem
\cite{bb84,bbbss92,ekert91,bbpssw96,dejmps96,bennett92,ms94,he94,ehpp94,%
lb96,qpa96b,qpa96a,bms96}.  Using novel techniques
\cite{yao95,mayers95a}, a proof of security against all attacks for
the quantum key distribution protocol of Bennett and Brassard was
obtained in 1996 \cite{mayers96}.  Related results were subsequently
obtained \cite{bm97,bbbgm98,fggnp97,gn97}, but as yet \cite{mayers96}
is the only known proof of security against all attacks.  A more
recent version of the proof with extension to the result is proposed
in \cite{mayers98}.  Also, the basic ideas of
\cite{ekert91,bbpssw96,qpa96b} might lead to a complete solution if we
accept fault tolerant computation (for example, see \cite{lc98}), but
this is not possible with current technology.

\begin{figure}
\setlength{\unitlength}{1mm}
\begin{center}
\begin{picture}(40,30)
\put(10,10){\vector(1,0){14}}
\put(25,10){ \bbB{}{0,2} }
\put(10,10){\vector(0,1){14}}
\put(10,25){ \bbB{}{1,2} }
\put(10,10){\vector(1,1){10}}
\put(21,18){\bbB{}{0,3}}
\put(10,10){\vector(-1,1){10}}
\put(-3,22){\bbB{}{1,3}}
\end{picture}
\caption{The BB84 states}
\label{bb84_picture}
\end{center}
\end{figure}
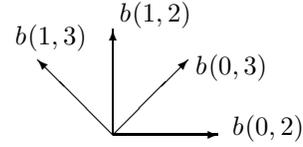

In the quantum transmission, Alice sends $n$ photons to Bob prepared
individually in one of the four BB84 states uniformly picked at
random.  The BB84 states denoted $\bbB{}{0,2}$, $\bbB{}{1,2}$,
$\bbB{}{0,3}$ and $\bbB{}{1,3}$ correspond to a photon polarized at
$0$, $90$, $45$ and $-45$ degrees respectively (see figure
\ref{bb84_picture}).  (We reserve the states $\bbB{}{0,1}$ and
$\bbB{}{1,1}$ for further use: we will have to add two other states in
our analysis.)
Bob measures each photon using either the rectilinear basis
$\{ \bbB{}{0,2}, \bbB{}{1,2} \}$ or the diagonal basis
$\{\bbB{}{0,3}, \bbB{}{1,3}\}$ uniformly chosen at random.

The basic idea of the protocol is the following.  Both, Eve and Bob,
do not know Alice's bases until after the quantum transmission.  Eve
cannot obtain information without creating a disturbance which can be
detected.  Bob also disturbs the state when he uses the wrong basis,
but this is not a problem. After the quantum transmission, Alice and
Bob announce their bases.  Alice and Bob share a bit when their bases
are identical, so they know which bits they share.  The key point is
that it's too late for Eve because the photons are on Bob's side.
However, the security of the protocol relies on the fact that the
source behaves as specified, and this is the main subject of this
paper.

Informally, the source used in the original BB84-protocol \cite{bb84}
can be described as a blackbox with two buttons on it: {\it
base2-button} and {\it base3-button}.  When Alice pushes the
base2-button, the output is either $(0, \bb{}{0,2} )$ or $(1,
\bb{}{1,2})$, where $\bb{}{0,2}$ and $\bb{}{1,2}$ form an orthonormal
basis of a two-dimensional system \Hb, with each possibility occurring
with probability $1/2$.  After the base\,$\alpha$-button is pushed, of
the output $(x,\bb{}{x,\alpha})$, only the vector $\bb{}{x,\alpha}$
goes out to Bob; bit $x$ is only visible to Alice.  Similarly, if
Alice pushes the base3-button, the output is either $(0, \bb{}{0,3})$
or $(1, \bb{}{1,3})$, with each possibility occurring with probability
$1/2$, where
\begin{eqnarray}
\bb{}{0,3} &=& (\bb{}{0,2} + \bb{}{1,2}) / \sqrt{2},\nonumber\\
\bb{}{1,3} &=& (-\bb{}{0,2} + \bb{}{1,2}) / \sqrt{2}.
\end{eqnarray}

The suggested way in \cite{bb84,bbbss92} to achieve the above is to
have the blackbox generates a fixed state, say $\bb{}{0,2}$, then the
bit $x \in \{0,1\}$ is uniformly chosen at random and this state is
rotated of an appropriate angle to create the desired state
$\bb{}{x,\alpha}$ (assuming that the base$\, \alpha$-button is
pressed).  The security proof of the protocol extends to sources
beyond mentioned above.  To obtain our self-testing source, we need to
consider a different type of sources.  A {\it conjugate coding source}
$S = (H_A \otimes H_B, \Psi , M_2, M_3)$ consists of a pure state
$\Psi$ in a Hilbert space $H_A \otimes H_B$, and two measurements
(each binary-valued) $M_2$, $M_3$ defined on $H_A \otimes H_B$ but
operating only on coordinates in $H_A$.  Pushing base2-button,
base3-button performs respectively measurement $M_2$, $M_3$.  (We have
restricted the form of the initial state to be a pure state
$\ket{\Psi}$ instead of a general mixed state.  This is without loss
of generality for our result, as we will see.)

Let $P_{\alpha}^+, P_{\alpha}^-$, where $\alpha \in \{ 2,3\}$, denote
the projection operators to the subspaces corresponding to the
outcomes $0,1$ for measurement $M_\alpha$.  (We sometimes use the
notation $P_{\alpha}^\pm$ to denote the measurement $M_\alpha$
itself.) After performing the measurement, only the coordinates in
$H_B$ are made available for transmission.  Thus, if button $\alpha$
is pushed with outcome $0$, the density operator in the transmitted
beam is $tr_A(P_{\alpha}^+ \ket{\Psi}\bra{\Psi} P_{\alpha}^+)$.  For
convenience, we sometimes identify $+$ with $0$, and $-$ with $1$.
Thus, if button $\alpha$ is pushed with outcome $x$, the density
operator is $tr_A(P_{\alpha}^x \ket{\Psi}\bra{\Psi} P_{\alpha}^x)$.

The security proof of the protocol is valid if the source satisfies, 
for $x \in \{0,1\}$, 
the conditions
\begin{eqnarray}
tr_A(P_{2}^x \ket{\Psi} \bra{\Psi}
P_{2}^x) &=& \ket{\bb{}{x,2}}\bra{\bb{}{x,2}}/2,\nonumber\\
tr_A(P_{3}^x  \ket{\Psi} \bra{\Psi}
P_{3}^x) &= & \ket{\bb{}{x,3}} \bra{\bb{}{x,3}}/2,
\end{eqnarray}
where $(\bb{}{0,2}, \bb{}{1,2})$ and
$(\bb{}{0,3}, \bb{}{1,3})$
are orthonormal bases that satisfy equation~(1).

It is well known (and easy to see) that the following source satisfies
the above condition.  Let $H_A$, $H_B$ each be a two-dimensional
Hilbert space.  Let $({\bbA{}{0,2}}, {\bbA{}{1,2}}), (\bbA{}{0,3},
\bbA{}{1,3})$ be two pairs of orthonormal bases of $H_A$ related by
equation (1); similarly let $({\bbB{}{0,2}}, {\bbB{}{1,2}}),
(\bbB{}{0,3}, \bbB{}{1,3})$ be two pairs of orthonormal bases related
by equation (1) for $H_B$.  Let $\Psi$ be the Bell state
$(\ket{\bbA{}{0,2}}\ket{\bbB{}{0,2}} + \ket{\bbA{}{1,2}}
\ket{\bbB{}{1,2}})/ \sqrt{2} = (\ket{\bbA{}{0,3}}\ket{\bbB{}{0,3}} +
\ket{\bbA{}{1,3}} \ket{\bbB{}{1,3}})/ \sqrt{2} $.  Let $M_2, M_3$ be
two measurements on $H_A \otimes H_B$ that operate only on the
coordinates in $H_A$.  The measurement $M_2$ consists of the two
orthogonal subspaces $\ket{\bbA{}{0,2}} \otimes H_B$,
$\ket{\bbA{}{1,2}} \otimes H_B$.  The measurement $M_3$ consists of
the two orthogonal subspaces $\ket{\bbA{}{0,3}} \otimes H_B$,
$\ket{\bbA{}{1,3}} \otimes H_B$.  If we restrict them to $H_A$ only,
the measurement $M_2$, $M_3$ are the measurements in the bases
$(\bbA{}{0,2}, \bbA{}{1,2})$ and $(\bbA{}{0,3}, \bbA{}{1,3})$
respectively.  Clearly, (2) is true.  We call this source the {\it
perfect system}.

More generally, the security proof extends to systems that behave like
a mixture of orthogonal ideal systems.  A source is an {\it extended
perfect system} if there exist in $H_B$ orthogonal two dimensional
subspaces $H_i$ ($i \in I$, some index set), with $\bb{i}{x,2}{},
\bb{i}{x,3}$ denoting states in $H_i$ that respect the same
ortogonality condition as the above states $\bbB{}{x,2}, \bbB{}{x,3}$
in $H_B$ and equation (1), such that for some probability distribution
$p_i$ on $i \in I$,
\begin{eqnarray}
tr_A(P_{2}^x   \ket{\Psi}\bra{\Psi}
P_{2}^x) &=& \sum_{i \in I} p_i \;  \ket{\bbB{i}{x,2}}
\bra{\bbB{i}{x,2}}/2, \nonumber\\
tr_A(P_{3}^x   \ket{\Psi}\bra{\Psi}
P_{3}^x) &= &\sum_{i \in I} p_i  \; \ket{\bbB{i}{x,2}}
\bra{\bbB{i}{x,3}}/2.
\end{eqnarray}

Now comes the question.  If a manufacturer hands over a source and
claims that it is a perfect system, how can we check this claims, or
at least, makes sure that it is an extended perfect system?

If the source is a perfect system, let $N_2, N_3$ be the measurements
operating on $H_B$ in exactly the same way as $M_2, M_3$ on $H_A$.
That is, let $R_\alpha^+, R_\alpha^-$ (where $\alpha \in \{ 2,3\}$) be
the projection operators to subspaces by $N_\alpha$ with outcome
$0,1$; $R_2^+, R_2^-$ project to $ H_A \otimes \ket{\bbB{}{0,2}}$,
$H_A \otimes \ket{\bbB{1,2}}$, and $R_3^+, R_3^-$ project to $ H_A
\otimes \ket{\bbB{}{0,3}}$, $H_A \otimes \ket{\bbB{}{1,3}}$,
respectively.  Now observe that the following are true for $\alpha
\not= \beta \in \{ 2,3\}$, $x, y \in \{+,-\}$,
\begin{eqnarray}
|| P_\alpha^x \ket{\Psi}||^2 &=& 1/2,\nonumber\\
{|| R_\alpha^y P_\alpha^x \ket{\Psi}||^2  \over || P_\alpha^x
\ket{\Psi}||^2 }&=&
\delta_{x,y},\nonumber\\
{|| R_\beta^y P_\alpha^x \ket{\Psi}||^2  \over || P_\alpha^x
\ket{\Psi}||^2 } 
&=&  1/2.
\end{eqnarray}
We can ask the manufacturer to provide in addition two measuring
devices outside the blackbox corresponding to $N_2,N_3$. A test can be
executed to verify that these equations are satisfied (see the related
discussion in the Introduction).  Furthermore, as a matter of physical
implementation, to make sure that $M's$ and $N's$ operate on $H_A$,
$H_B$ respectively, we can further demand that the buttons are
replaced by two measuring devices outside the blackbox.  Is that
sufficient to guarantee that we have at least an extended perfect
system?

Unfortunately, the answer is NO.  It is not hard to construct examples
where (4) is satisfied, but it is not an extended ideal system (and in
fact, security is gravely compromised).

However, as we will see, if we add one more measurement appropriately
on each side, and perform the corresponding checks, then it gurantees
to be an extended perfect system. That will be the main result of this
paper.

\section{Main Theorem}
An object $S=(H_A \otimes H_B, \ket{\Psi}, P^\pm_1,
P^\pm_2, P^\pm_3, R^\pm_1, R^\pm_2, R^\pm_3)$ is called an 
{\it ideal source} if the following are valid: each of $H_A, H_B$ is a
2-dimensional Hilbert space with
$(\bbA{}{0,2},\bbA{}{1,2}),(\bbA{}{0,3},\bbA{}{1,3})$ being a pair of
orthonormal basis of $H_A$ satisfing equation (1), and
$(\bbB{}{0,2},\bbB{}{1,2}),(\bbB{}{0,3},\bbB{}{1,3})$ being a pair of
orthonormal basis of $H_B$ satisfying equation (1); $\Psi$ is the Bell
state $(\bbA{}{0,\alpha}\bbB{}{0,\alpha} +
\bbA{}{1,\alpha}\bbB{}{1,\alpha})/ \sqrt{2}$; $P_2^+, P_2^- $ are the
projection operators on the states $\bbA{}{0,2}$, $\bbA{}{1,2}$
respectively; $P_3^+, P_3^- $ are the projection operators on the
states $\bbA{}{0,3}$, $\bbA{}{1,3}$ respectively.  To describe
$P_1^\pm$, let $\bbA{}{x,1}$ ($x \in \{0,1\}$) be the state
$\bbA{}{x,2} + \bbA{}{x,3}$ after being normalized to unit length. The
states $\bbA{}{x,1}$ and $\bbB{}{x,1}$ have a particular status in
our proof, and we alternatively denote $\bbA{}{x,1} = \ket{x}$ and
$\bbB{}{x,1} = \ket{x}'$.  Then $P_1^+, P_1^-$ are respectively the
projection operators on the states $\ket{0}$, $\ket{1}$.  As usual, we
consider $P^x_\alpha$ and $P^x_\alpha \otimes {\bf I}$ as two
alternative notations for one and the same projection operators on
$\Ha \otimes \Hb$.  Clearly, $P_1^\pm, P_2^\pm, P_3^\pm$ are the
projection operators on $\Ha \otimes \Hb$ corresponding to measuring
$H_A$ with respect to three bases of $H_A$ (the bases for $P_2^\pm$,
$P_3^\pm$ at an angle of $-\pi/8$, $+\pi/8$ with repect to the basis
for $P_1^\pm$).

The projection operators $R_1^\pm, R_2^\pm, R_3^\pm$
operate on coordinates in $H_B$, and are similarly
defined as the $P's$.  Let
\begin{eqnarray}
p_{\alpha,\beta}(x,y) &=& || R_\alpha^x P_\beta^y\ket{\Psi}||^2.
\end{eqnarray}
These numbers can be easily computed.  For example,
$p_{1,2}(0,0)= (\cos (\pi/8))^2/2$ and $p_{1,2}(0,1)= (\sin (\pi/8))^2/2$.

A  {\it self-checking source}\hfill\break 
$S=(H_A \otimes H_B, \ket{\Psi}, P^\pm_1,
P^\pm_2, P^\pm_3, R^\pm_1, R^\pm_2, R^\pm_3)$  consists of an
initial state $\ket{\Psi} \in H_A \otimes H_B$, three measurements
$P^\pm_1,
P^\pm_2, P^\pm_3$ acting on coordinates in $H_A$,
and  three measurements
$R^\pm_1,
R^\pm_2, R^\pm_3$ acting on coordinates in $H_B$,
such that the following conditions are satisfied:
\begin{eqnarray}
|| R_\alpha^x P_\beta^y\ket{\Psi}||^2
&=& p_{\alpha,\beta}(x,y) .
\end{eqnarray}

We will see that a self-checking source gives
rise to an extended ideal system.

An {\it extended ideal source}\hfill\break
$S=(H_A \otimes H_B, \ket{\Psi}, P^\pm_1,
P^\pm_2, P^\pm_3, R^\pm_1, R^\pm_2, R^\pm_3)$ is an
orthogonal sum of ideal sources in
a similar sense as an extended perfect
system in relation to perfect systems.
That is, if there is an index set $I$,
orthogonal
two dimensional subspaces
$K_i \subseteq H_A$
with $\bbA{i}{x}$ (or alternatively $\bbA{i}{x,1}$) denoting the
state $\ket{x}$ in $K_i$,
orthogonal
two dimensional subspaces
$H_i \subseteq H_B$
with $\bbB{i}{x,\alpha}$ (or alternatively $\bbB{i}{x,1}$) denoting the
state $\ket{x}'$ in $H_i$, such
that for some (possibly complex) numbers
$\alpha_i$ on $i \in I$ with $\sum_{i \in I} |\alpha|^2 =1$,
\begin{eqnarray*}
\Psi &=& \sum_{i \in I} \alpha_i \; (\bbA{i}{0} \otimes \bbB{i}{0}
              +    \bbA{i}{1} \otimes \bbB{i}{1}  ).
\end{eqnarray*}

Furthermore, for each $i$, for every projection $P \in \{P_1^\pm,
P_2^\pm, P_3^\pm\}$, $P$ acts exactly on $K_i$ like the corresponding
projection on $H_A$ in the ideal source case.  That is, if $P
|x\rangle = \lambda_0 |0\rangle + \lambda_1 |1\rangle$ in the ideal
case, we have that $P \bbA{i}{x} = \lambda_0 \bbA{i}{0} + \lambda_1
\bbA{i}{1}$.  The following fact is easy to verify.

\noindent{\large\bf Fact 1} Any extended ideal source
is a self-checking source.

Also, it is clear that from any self-checking source,
by omitting the measurements $P_1^\pm, R_1^\pm$,
one obtains a conjugate coding source.

\noindent{\large\bf Fact 2} The conjugate coding
source obtained from an extended ideal
source must be an extended perfect system.

The converse of fact 1 is our main theorem.
\vskip11pt
\noindent{\large\bf Main Theorem} Any self-checking
source is an extended ideal source.
\vskip11pt
It follows from the Main Theorem and Fact 2
that a self-checking source provides
an adequate source for the BB84 quantum key
distribution protocol~\cite{bb84,bbbss92}.

We remark that in our definition of self-checking
source, the restriction of the initial state to
a pure state $\ket{\Psi}$ instead of a mixed state
$\rho$ is not a real restriction.  Given
a source with a mixed
state $\rho$ satisfying equation (6), we can construct one with a pure
state $\ket{\Psi}$ (by enlarging appropriately $H_A$) satisfying
(6).  We can apply the Main Theorem to this new source,
and conclude that it also gives rise to an adequate source
for the BB84-protocol.

It is well known, from discussions about
{\it EPR Experiments} (see e.g. \cite{ag86}),
that quantities such as
$|| R_\alpha^x P_\beta^y\ket{\Psi}||^2 $
exhibit behavior characteristic of quantum
systems that cannot be explained by classical
theories.  One may
view our main result as stating that
such constraints are sometimes strong enough
to yield precise structural information about
the given quantum system; in this case it
has to be an orthogonal sum of EPR pairs

\section{Proof of Main Theorem}

We give in this Section a sketch of the main steps in the proof. 
Let $S=(H_A \otimes H_B, \ket{\Psi}, P^\pm_1,
P^\pm_2, P^\pm_3, R^\pm_1, R^\pm_2, R^\pm_3)$  
be a self-checking source.  We show that 
it must be an extended ideal source.
 
In Section 4.1, we derive some structural properties 
of the projection operators as imposed by 
the self-checking conditions, but without 
considering in details the constraints due to 
the tensor product nature of the state space. 
In Sections 4.2 and 4.3, the state is decomposed 
explicitly in terms of tensor products, 
and the properties derived in Section 4.1 
are used to show that this decomposition 
satisfies the conditions stated in 
the Main Theorem.

\subsection{Properties of Projections}

In this subsection, we present some properties of 
the projected states (such as 
$P_1^+ \Psi, P_1^+ R_2^- \Psi$) as  
consequences of the constraints put on 
self-checking sources.  
The proofs of these lemmas are somewhat lengthy, 
and will be left to the complete paper.

\noindent{\bf Lemma 1}  
For every $\alpha \in \{1,2,3\}$
and $x \in \{+,-\}$, we have $P^x_\alpha \Psi = R^x_\alpha \Psi$.

Let $v_i \in V, w_i \in W$ 
for $1 \leq i \leq m$, where 
$V, W$ are two Hilbert spaces.  We say that  
$(v_1, v_2, \cdots, v_m)$ is {\it isormorphic} to $(w_1, w_2, \cdots, w_m)$ 
if there is an inner-product-preserving 
linear mapping $f: V \rightarrow W$ such that 
$w_i = f(v_i)$ for all $i$.

Let $\theta = \pi / 8$,  
and 
$u_1,u_2, \cdots, u_5$ be elements of ${\boldmath C}^2$ defined 
by
\begin{eqnarray*}
u_1 &=& (1,0),\\
u_2 &=& (\cos^2 \theta, \sin \theta \cos \theta),\\
u_3 &=& (\sin^2 \theta, -\sin \theta \cos \theta ),\\
u_4 &=& (\cos^2 \theta, - \sin \theta \cos \theta),\\
u_5 &=& (\sin^2 \theta, \sin \theta \cos \theta ).
\end{eqnarray*}

\noindent {\bf Lemma 2} $(u_1, u_2, \cdots, u_5)$ is isomorphic to 
$\sqrt{2}(P_1^+\Psi, P_1^+R_3^+ \Psi, P_1^+R_3^- \Psi, 
P_1^+R_2^+ \Psi , P_1^+R_2^- \Psi)$.
\vskip15pt

\noindent {\bf Lemma 3} Let $h = P_1^+R_3^+ \Psi - P_1^+ R_2^+ \Psi$. 
Then $R_1^- h = h$.
\vskip15pt

\noindent {\bf Lemma 4} Let $k = P_2^+P_1^+ \Psi - (\cos \theta)^2P_1^+ \Psi$. 
Then $(R_2^+ - R_3^+) k = 2  (\sin \theta \cos \theta)^2 P_1^- \Psi$.
\vskip15pt

Since there is a symmetry between the 
projection operators $P$ and $R$, the following is clearly true.
\vskip13pt
\noindent {\bf Lemma 5} Lemmas 2-4 remain valid if the projection 
operators $P$ and $R$ are exchanged.

\vskip15pt

\subsection{The Decomposition}

We now prove that the state $\Psi \in H_A \otimes H_B$ can be 
decomposed into the direct 
sum of EPR pairs. We begin with a decomposition 
of $P_1^+ \Psi$, which is equal to 
$R_1^+ \Psi$ by Lemma 1.

\noindent{\bf Lemma 6} One can write 
\[  P_1^+ \Psi = \sum_{i \in I} \alpha_i a_i(0) \otimes b_i(0) \]
where  $I$ is an index set, $\alpha_i$ are complex numbers, and 
$a_i(0) \in H_A (i \in I)$, $b_i(0) \in H_B (i \in I)$ are two 
respectively orthonormal sets of eigenvectors of 
the operators $P_1^+$ (acting on $H_A$) and 
$R_1^+$ (acting on $H_B$).

\noindent{\it Proof} The lemma is  
proved with the help of Schmidt decomposition theorem 
\cite{schmidt06} \cite{hjw93}. 
We omit the details here. $\Box$

Let $\beta = (2 \sin \theta cos \theta)^{-1}$.  
Define $a_i(1) = \beta (P_3^+ - P_2^+) a_i(0)$, 
and $b_i(1) = \beta (R_3^+ - R_2^+) b_i(0)$ for 
$i \in I$.  Let $K_i \subseteq H_A$ be the 
subspace spanned by $a_i(0)$ and $a_i(1)$; 
Let $H_i \subseteq H_B$ be the 
subspace spanned by $b_i(0)$ and $b_i(1)$.  
The plan is to show that 
\[
\Psi = \sum_{i \in I} \alpha_i (\bbA{i}{0} \otimes \bbB{i}{0} + \bbA{i}{1} \otimes
\bbB{i}{1}),
\]
and that $K_i, H_i$ have all the 
properties required to satisfy the Main Theorem.

In the remainder of this subsection, we use Lemmas 2-5 to 
show that 
each $H_i$ ($K_i$) behaves correctly under 
the projection operators $R_{\gamma}^x$ ($P_{\gamma}^x$). 
In the next subsection, we complete the proof by 
showing that all $H_i$ ($K_i$) are orthogonal to each other.

By Lemma 2, $(u_1, u_2, \cdots, u_5)$ is isomorphic to 
$\sqrt{2}
(P_1^+\Psi, P_1^+R_3^+ \Psi, P_1^+R_3^- \Psi, P_1^+R_2^+ \Psi , 
P_1^+R_2^- \Psi)$. In particular, this implies that any 
linear relation $\sum_j \lambda_j u_j = 0$ must also be 
satisfied if $u_j$ are replaced by the appropriate projected states. 
Now
\begin{eqnarray*}
P_1^+\Psi &=&  \sum_{i \in I} \alpha_i a_i(0) \otimes b_i(0),\\
P_1^+R_3^+ \Psi &=&  \sum_{i \in I} \alpha_i a_i(0) \otimes  R_3^+b_i(0),\\ 
P_1^+R_3^- \Psi &=&  \sum_{i \in I} \alpha_i a_i(0) \otimes R_3^-b_i(0),\\
P_1^+R_2^+ \Psi  &=&  \sum_{i \in I} \alpha_i a_i(0) \otimes R_2^+ b_i(0),\\
P_1^+R_2^- \Psi &=&  \sum_{i \in I} \alpha_i a_i(0) \otimes R_2^- b_i(0).
\end{eqnarray*}
This means that, for each $i \in I$, any 
linear relation $\sum_j \lambda_j u_j = 0$ must also be 
satisfied if we make the following substitutions:
\begin{eqnarray*}
u_1 \leftarrow b_i(0),\\
u_2 \leftarrow R_3^+b_i(0),\\
u_3 \leftarrow R_3^-b_i(0),\\
u_4 \leftarrow R_2^+b_i(0),\\
u_5 \leftarrow R_2^- b_i(0).
\end{eqnarray*}

\noindent{\bf Lemma 7}  
For each $i \in I$, $(u_1, u_2, \cdots, u_5)$ is isomorphic to 
$(b_i(0), R_3^+b_i(0), R_3^-b_i(0), R_2^+b_i(0), R_2^-b_i(0))$.
\vskip 11pt
\noindent{\it Proof} Use the preceding observation and 
the orthogonality between $R_3^+b_i(0)$ and $ R_3^-b_i(0)$, 
and the orthogonality between  
$R_2^+b_i(0)$ and $R_2^-b_i(0))$. 
We omit the details here. $\Box$
\vskip12pt

Note that $b_i(1) = \beta (R_3^+ - R_2^+) b_i(0)$ by definition. 
>From Lemma 7, it is easy to see that $b_i(1)$ is a unit vector 
perpendicular to $b_i(0)$.  In fact, $b_i(1)$ is mapped 
to the vector $(0,1)$ under the isomorphism in Lemma 7.

>From Lemma 7, for the purpose of vectors in the 
space $H_i$, the projection operators $R_3^+, R_3^-$ correspond 
to choosing the coordinate system obtained from 
the system $(b_i(0), b_i(1))$ rotated by the angle $\theta$;  
similarly, $R_2^+, R_2^-$ correspond 
to choosing a coordinate system 
obtained from 
the system $(b_i(0), b_i(1))$ 
rotated by the angle $-\theta$. 
It remains to show that $R_1^+, R_1^-$ correspond to 
the coordinate system $(b_i(0), b_i(1))$ itself. 
By definition $R_1^+ b_i(0) = b_i(0)$. It remains 
to prove that $R_1^- b_i(1) = b_i(1)$. 

To do that, we use Lemma 3.  Observe that 
\begin{eqnarray*}
h &=& P_1^+R_3^- \Psi - P_1^+ R_2^- \Psi\\
  &=&  \sum_{i \in I} \alpha_i a_i(0) \otimes  R_3^+b_i(0)\\ 
 &&  \ \ \ - \sum_{i \in I} \alpha_i a_i(0) \otimes  R_2^+b_i(0)\\ 
&=&  \sum_{i \in I} \alpha_i a_i(0) \otimes  (R_3^+ - R_2^+)b_i(0)\\ 
&=& \beta^{-1} \sum_{i \in I} \alpha_i a_i(0) \otimes  b_i(1).
\end{eqnarray*}
Since $R_1^- h = h$ by Lemma 3, we must have $R_1^- b_i(1) = b_i(1)$. 
This completes the proof that the projection operators $R_\gamma^x$ 
behave as required on the subspace $H_i$.

As stated explicitly in Lemma 5, we can obtain the symmetric 
statement that the the projection operators $P_\gamma^x$ 
behave as required on the subspace $K_i$.

Now that we have determined the behavior of the projection 
operators on $K_i, H_i$, we can in principle calculate 
any polynomial of the projection operators on the 
state $P_1^+ \Psi$.  By Lemma 4, $P_1^- \Psi$ can be 
written as
\[ P_1^- \Psi = 2 \beta^2 (R_2^+ - R_3^+)(P_2^+ - 
\cos^2 \theta )P_1^+ \Psi.\]
This gives
\[ P_1^- \Psi = 2 \beta^2 \sum_{i \in I} \alpha_i 
(P_2^+ - \cos^2 \theta ) a_i(0) \otimes 
(R_2^+ - R_3^+) b_i(0).\]
After applying the rules and symplifying, we 
obtain 
\[  P_1^- \Psi =  \sum_{i \in I} \alpha_i 
a_i(1) \otimes  b_i(1).\]
As $\Psi = P_1^+ \Psi + P_1^- \Psi$, this proves
\[
\Psi = \sum_{i \in I} \alpha_i (\bbA{i}{0} \otimes \bbB{i}{0} + \bbA{i}{1} \otimes
\bbB{i}{1}).
\]

\subsection{Completing the Proof}

It remains to show that all $H_i$ are orthogonal to 
each other.  (A symmetric argument then 
shows that all $K_i$ are also orthogonal 
to each other.) 

Let $i \not= j \in I$.  Assume that $H_i$ is not orthogonal 
to $H_j$.  We derive a contradiction.  By definition, 
$H_i$ is spanned by $b_i(0), b_i(1)$, and 
$H_j$ is spanned by $b_j(0), b_j(1)$.  
Clearly, $b_i(1)$ and $b_j(1)$ are not orthogonal 
to each other, as all the other pairs $(b_i(x), b_j(y))$ 
are orthogonal.

Choose a coordinate system for the space spanned by 
the four vectors such that
\begin{eqnarray*}
b_i(0) &=& (1, 0, 0, 0),\\
b_i(1) &=& (0, 1, 0, 0),\\
b_j(0) &=& (0, 0, 1, 0),\\
b_j(1) &=& (0, s, 0, t),
\end{eqnarray*}
where $s \not= 0$.  From our knowledge about 
the behavior of $R_3$, we infer that 
$R_3^+ b_j(0) = (\cos \theta )w$ 
where $w = \cos \theta b_j(0) + \sin \theta b_j(1)
= (0, s \sin \theta , \cos \theta, t \sin \theta)$. 
Similary, $R_3^- b_i(0) = (\sin \theta)w'$ 
where $w' = -\sin \theta b_i(0) + \cos \theta b_i(1)
= (- \sin \theta , \cos \theta , 0, 0)$.  As the 
inner product of $w$ and $w'$ is $s \sin \theta \cos \theta$ 
which is non-zero, we conclude that $R_3^+ b_j(0)$ and $R_3^- b_i(0)$ 
are not orthogonal. This contradicts 
the fact that $R_3^+, R_3^-$ are 
projection operators to orthogonal subspaces. 
This completes the proof.

\section{Concluding Remarks}

The security problem for imperfect source is a
difficult one to deal with.  The present paper
is a step in only one possible direction.  We have
also limited ourselves to the simplist case when
the correlation probabilities $p_{\alpha, \beta}(x,y)$
are assumed to be measurable precisely.  We leave open as
future research topics for extensions to more general models.

\end{document}